\documentclass[twocolumn,a4paper,preprintnumbers,showkeys,aps,pra]{revtex4}
\usepackage{graphicx, epstopdf, natbib, doi, float, amsmath, amssymb, color, mathtools}
\graphicspath{images/}
\DeclarePairedDelimiter\abs{\lvert}{\rvert}
\DeclarePairedDelimiter\norm{\lVert}{\rVert}
\makeatletter
\let\oldabs\abs
\def\abs{\@ifstar{\oldabs}{\oldabs*}}
\let\oldnorm\norm
\def\norm{\@ifstar{\oldnorm}{\oldnorm*}}
\makeatother

\begin{document}
\title{Tunable multiwindow magnomechanically induced transparency, Fano resonances, and slow to fast light conversion} 
\author{Kamran Ullah}
\email[Electronic address:\ ]{kamran@phys.qau.edu.pk}
\author{M. Tahir Naseem}
\email[Electronic address:\ ]{mnaseem16@ku.edu.tr}
\author{\"{O}zg\"{u}r E.\ M\"{u}stecapl{\i}o\u{g}lu}
\email[Electronic address:\ ]{omustecap@ku.edu.tr}
\affiliation{Department of Physics, Ko\c{c} University, Sar{\i}yer, \.Istanbul, 34450, Turkey}
\begin{abstract}
We investigate the absorption and transmission properties of a weak probe field under the influence of a strong
control field in a cavity magnomechanical system. The system consists of two ferromagnetic-material yttrium
iron garnet (YIG) spheres coupled to a single cavity mode. In addition to two magnon-induced transparencies
(MITs) that arise due to magnon-photon interactions, we observe a magnomechanically induced transparency
(MMIT) due to the presence of nonlinear magnon-phonon interaction. We discuss the emergence of Fano
resonances and explain the splitting of a single Fano profile to double and triple Fano profiles due to additional
couplings in the proposed system. Moreover, by considering a two-YIG system, the group delay of the probe
field can be enhanced by one order of magnitude as compared with a single-YIG magnomechanical system.
Furthermore, we show that the group delay depends on the tunability of the coupling strength of the first YIG
with respect to the coupling frequency of the second YIG, and vice versa. This helps to achieve larger group
delays for weak magnon-photon coupling strengths.

\end{abstract}
\maketitle
\section{introduction}

Storing information in different frequency modes of light has attracted much attention due to its critical role in high-speed, long-distance  quantum communication applications~\cite{PhysRevA.79.052329, PhysRevLett.113.053603, PhysRevA.93.032327}. The spectral distinction  of optical signals eliminates their unintentional coupling to the stationary information or memory nodes in a communication network. For that aim, multiple transparency window Electromagnetically Induced Transparency (EIT) schemes have been considered for multiband quantum memory implementations mainly in the medium of three-level cold atoms. Experimental demonstrations of three EIT windows have been reported~\cite{PhysRevA.62.053802}, and extended to seven windows using external fields~\cite{doi:10.1080/09500340.2014.904019}. Observation of nine EIT windows has been experimentally demonstrated quite recently, using an external magnetic field in a vapor cell of Rubidium atoms~\cite{BHUSHAN2019125885}. A practical question is if such results can be achieved at higher temperatures, for example, for a room temperature multiband quantum memory.

In recent years, remarkable developments have been achieved to strongly couple spin ensembles to cavity photons, leading to the emerging field of cavity spintronics. Quanta of spin waves, magnons, are highly robust against temperature~\cite{zhang2015cavity, PhysRevLett.113.083603, PhysRevApplied.2.054002, PhysRevLett.111.127003, PhysRevLett.113.156401}, and hence significant magnon-photon hybridization and magnetically induced transparency (MIT) have been successfully demonstrated even at room temperature~\cite{PhysRevLett.113.156401}. Tunable slow light and its conversion to fast light based upon room temperature MIT has been theoretically shown recently~\cite{8701447}. Besides, at strong magnon-photon interaction, a wide tunability of slow light via applied magnetic field has been shown in~\cite{Kong:19}. These results demonstrate the promising value of these systems for practical quantum memories ~\cite{8701447}. Here we explore how to split such a MIT window into multiple bands for a room temperature multimode quantum memory. Our idea is to exploit the coupling of magnons to thermal vibrations, which is known to yield magnomechanically induced transparency (MMIT)~\cite{Zhange1501286}, in combination with multiple spin ensembles to achieve multiple bands in MIT. We also discuss the emergence of Fano resonance in the output spectrum and explore the suitable system parameters for its observation. Fano resonance was first reported in the atomic systems~\cite{PhysRev.124.1866}, and it emerges due to the quantum interference of different transition amplitudes which give minima in the absorption profile. In later years, it has been discussed in different physical systems, such as photonic crystal~\cite{PhysRevLett.103.023901}, coupled microresonators~\cite{PhysRevA.82.065804}, optomechanical system~\cite{Kaur_2016}. Recently, Fano-like asymmetric shapes have been experimentally reported in a hybrid cavity magnomechanical system~\cite{Zhange1501286}.

Our model consists of two ferromagnetic insulators, specifically yttrium iron garnets (YIGs), hosting long-lived magnons at room temperature, placed inside a three-dimensional (3D) microwave cavity; we remark that another equivalent embodiment of our model could be to place the YIGs on top of a superconducting co-planar waveguide, which can have further practical significance being an on-chip device~\cite{PhysRevB.93.174427}. Specific benefits of YIG as the host of spin ensemble over other systems, such as paramagnetic spin ensembles in nitrogen-vacancy centers is due to its high spin density of $2.1\times10^{22}$ $\mu$B  $cm^{-3}$ ($\mu$B  is the Bohr magneton) and high room temperature spin polarization below the Curie temperature (559 K). In addition to multimode quantum memories, our results can be directly advantageous for readily integrated microwave circuit applications at room temperature such as multimode quantum transducers coupling different systems at different frequencies~\cite{doi:10.1063/1.4945685}, tunable frequency quantum sensors~\cite{huang2014self} or fast light enhanced gyroscopes~\cite{PhysRev.73.155}. In addition to the magnetic dipole interaction between the cavity field and the spin ensemble, we take into account coupling between the magnons and the quanta of YIG lattice vibrations, phonons, arising due to the magnetostrictive force~\cite{Zhange1501286}. We 
only consider the Kittel mode~\cite{PhysRevLett.121.203601} of the ferromagnetic resonance modes of the magnons. Such three-body quantum systems can be of fundamental significance to examine macroscopic quantum phenomena towards thermodynamic limit and quantum to classical transitions~\cite{Li_2019}. 
  
In our model, tunable slow and fast light emerges as a natural consequence of tunable splitting of MIT window. Slow-light propagation at room temperature has been investigated recently in a cavity-magnon system and the group delays are found to be in the $\sim\mu$s range~\cite{8701447}. In a single YIG magnomechanical system with strong magnon-photon coupling strength, slow-light has achieved with a maximum group delay of $<0.8$ ms~\cite{Kong:19}. In this paper, we discuss the slow and fast light in a two YIGs magnomechanical system. Further, we exlain the group delay depends on the tunability of the magnon-photon coupling of the first YIG (YIG1) with respect to magnon-photon coupling of the second YIG (YIG2). This not only helps to achieve larger group delays at weak magnon-photon coupling, but also increase the group delay of the transmitted probe field by one order of magnitude, which is not possible with a single YIG system~\cite{Kong:19}.

The rest of the paper is organized as follows:
We describe the model system in Sec.~\ref{sec:model} and present dynamical equations with steady-state solutions. The results and discussions for MMIT are presented in the Sec.~\ref{sec:MMITwind}. We discuss the emergence and tunability of the multiple Fano resonances in Sec.~\ref{sec:fano}.
Next, in Sec.~\ref{sec:SlowFast}, we present the transmission of the probe field and discuss the group delays for slow and fast light propagation. Finally, in Sec.~\ref{sec:concul}, we present the conclusion of our work.
\begin{figure}
\centering
\includegraphics[width=9cm,height=5.5cm]{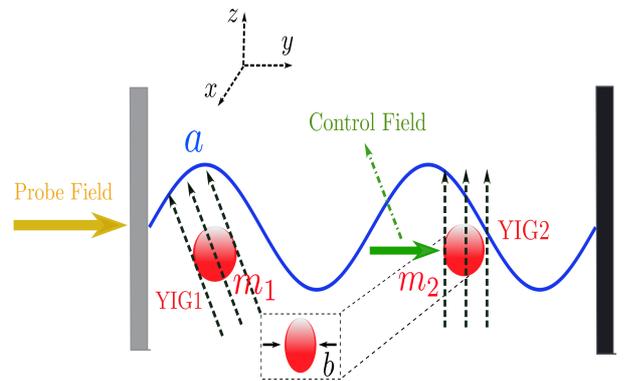}\label{}
\caption{(color online)~A schematic illustration of a hybrid cavity magnomechanical system. 
 It consists of two ferromagnetic yttrium iron garnet (YIG) spheres placed inside a microwave cavity. A Bias magnetic field is applied in the $z$ direction on each sphere, which excites the magnon modes, and these modes are strongly coupled with the cavity field. The bias magnetic fields activate the magnetostrictive (magnon-phonon) interaction in both YIGs. The single-magnon magnomechanical coupling strength is very weak~\cite{Zhange1501286}, and it depends on the spheres diameters and external bias field directions. Either by considering a larger YIG1 sphere or adjusting the direction of the bias field on it, the magnomechanical coupling of this sphere can be ignored. Here, we assume the direction of the bias field on YIG1 such that the single-magnon magnomechanical interaction becomes very weak and can be ignored~\cite{Zhange1501286}. However, the magnomechanical interaction of YIG2 is enhanced by directly driving its magnon mode via a microwave drive (y direction). 
This microwave drive plays the role of a control field in our model. Cavity, phonon, magnon modes are labeled as $a$, $b$, and $m_{i}$ ($i=1, 2$), respectively.} \label{fig:model}
\end{figure}

\section{System Hamiltonian and Theory}\label{sec:model}

We consider a hybrid cavity magnomechanical system that consists of two YIG spheres placed inside a microwave cavity, as shown in Fig.~\ref{fig:model}. A uniform bias magnetic field (z-direction) is applied on each sphere, which excites the magnon modes and these modes are coupled with the cavity field via magnetic dipole interaction.
The excitation of the magnon modes inside the spheres leads to the variation magnetization that results in the deformation of their lattice structures. The magnetostrictive force causes vibrations of the YIGs which establishes magnon-phonon interaction in these spheres.
The single-magnon magnomechanical coupling strength is very weak~\cite{Zhange1501286}, and it depends on the spheres diameters and external bias field directions. Either by considering a larger YIG1 sphere or adjusting the direction of the bias magnetic field on it, the magnomechanical coupling of this sphere can be ignored~\cite{Li_2019}. Here, we assume the direction of the bias field on YIG1 such that the single-magnon magnomechanical interaction becomes very weak and can be ignored~\cite{Zhange1501286}. However, the magnomechanical interaction of YIG2 is enhanced by directly driving its magnon mode via a external microwave drive. This microwave drive plays the role of a control field in our model. In addition, the cavity is driven by a weak probe field.\\
In this work, we consider high quality YIG spheres, each has a 250 $\mu$m diameter, and composed of ferric ions Fe$^{+3}$ of density $\rho=4.22\times10^{27} m^{-3}$. This causes a total spin $S=5/2 \rho V_{m}=7.07\times 10^{14}$, where $V_{m}$ is the volume of the YIG and $S$ is the collective spin operator which satisfy the algebra; $[S_{\alpha}, S_{\beta}]=i\varepsilon^{\alpha\beta\gamma}S_{\gamma}$. The Hamiltonian of the system reads~\cite{Li_2019}
\begin{equation}\label{eq:hamil}
\begin{split}
H/\hbar&=\omega_{a}\hat{a}^{\dagger}\hat{a}+\omega_{b}\hat{b}^{\dagger}\hat{b}+\sum_{j=1}^{2}[\omega_{j}{\hat{m}}_{j}^{\dag}{\hat{m}}_{j}+g_{j}({\hat{m}_{j}}^{\dag}\hat{a}+m_{j}\hat{a}^{\dag})]\\&+g_{mb}\hat{m}^{\dagger}_2\hat{m}_{2}(\hat{b}+\hat{b}^{\dagger})+i(\Omega_{d}\hat{m}^{\dagger}_{2}e^{-i\omega_{d}t}-\Omega_{d}^{\star}\hat{m}_{2}e^{i\omega_{d}t})\\&+i(\hat{a}^{\dagger}\varepsilon_{p}e^{-i\omega_{p}t}-\hat{a}\varepsilon_{p}^{\star}e^{i\omega_{p}t})
\end{split}
\end{equation}
where $a^{\dagger}(a)$ and $b^{\dagger}(b)$ are the creation (annihilation) operators of the cavity and phonon modes, respectively. The resonance frequencies of the cavity, phonon and magnon modes are denoted by $\omega_{a}$, $\omega_{b}$ and $\omega_{j}$, respectively. Moreover, $m_{j}$ is the bosonic operator of the Kittle mode of frequency $\omega_{j}$ and its coupling strength with the cavity mode is given by $g_{j}$.
The frequency $\omega_{j}$ of the magnon mode $m_{j}$ can be determined by using gyromagnetic ratio $\gamma_{j}$ and external bias magnetic field $H_{j}$ i.e., $\omega_{j}=\gamma_{j} H_{j}$ with $\gamma_{j}/2\pi=28$ GHz. The Rabi frequency $\Omega_{d}=\sqrt{5}/4\gamma \sqrt{N}B_{0}$~\cite{PhysRevLett.121.203601}, represents the coupling strength of the drive field of amplitude $B_{0}$ and frequency $\omega_{d}$. Furthermore, in Eq.~\eqref{eq:hamil}, $\omega_{p}$ is the probe field frequency having amplitude $\varepsilon_{p}$ which can be expressed as; $\varepsilon_{p}=\sqrt{2P_p\kappa_a/\hbar\omega_{p}}$.

 Note that in Eq.~\eqref{eq:hamil}, we have ignored the non-linear term $K\hat{m}^{\dagger}_{j}\hat{m}^{\dagger}_{j}\hat{m}_{j}\hat{m}_{j}$ that may arise due to strongly driven magnon mode~\cite{PhysRevLett.120.057202, PhysRevB.94.224410}. To ignore this nonlinear term, we must have $K\abs{\langle m_{2}\rangle}^3\ll\Omega$, and for the system parameters we consider in this work, this condition always satisfies. The Hamiltonian in Eq.~\eqref{eq:hamil} is written after applying the rotating-wave approximation in which fast oscillating terms $g_{j}(\hat{a}\hat{m}_{j}+\hat{a}^{\dagger}\hat{m}^{\dagger})$  are dropped. This is valid for $\omega_{a}, \omega_{j}\gg g_{j},\kappa_{a}, \kappa_{mj}$ which is the case to be considered in the present work.   
Where $\kappa_{a}$ and $\kappa_{mj}$ are the decay rates of the cavity and magnon modes, respectively.
In the frame rotating at the drive frequency $\omega_{d}$, the Hamiltonian of the system is given by
\begin{equation}\label{eq:hamil1}
\begin{split}
H/\hbar=&\Delta_{a}\hat{a}^{\dagger}\hat{a}+\omega_{b}\hat{b}^{\dagger}\hat{b}+\sum_{j=1}^{2}[\Delta_{mj}{\hat{m}}_{j}^{\dag}{\hat{m}}_{j}+g_{j}({\hat{m}}_{j}^{\dag}\hat{a}+\\&m_{j}\hat{a}^{\dag})]+g_{mb}\hat{m}^{\dagger}_{2}\hat{m}_{2}(\hat{b}+\hat{b}^{\dagger})+i(\Omega_{d}\hat{m}^{\dagger}_{2}-\Omega_{d}^{\star}\hat{m}_{2})+\\&i(\hat{a}^{\dagger}\varepsilon_{p}e^{-i\delta t}-\hat{a}\varepsilon^{\star}_{p}e^{i\delta t}),
\end{split}
\end{equation}
here, $\Delta_{a}=\omega_{a}-\omega_{d}$, $\Delta_{mj}=\omega_{j}-\omega_{d}$, and $\delta=\omega_{p}-\omega_{d}$. The quantum Heisenberg-Langevin equations based on the Hamiltonian in Eq.~\eqref{eq:hamil1} can be written as
\begin{equation}\label{eq:lang}
\begin{split}
&\dot{\hat{a}}=-i\Delta_{a}\hat{a}-i\sum_{j=1}^{2}g_{j}\hat{m}_{j}-\kappa_{a}\hat{a}+\varepsilon_{p}e^{-i\delta t}+\sqrt{2\kappa_{a}}\hat{a}^{in}(t), \\   
&\dot{\hat{b}}=-i\omega_{b}\hat{b}-ig_{mb}{\hat{m}}^{\dagger}_{2}\hat{m}_{2}-\kappa_{b}\hat{b}+\sqrt{2\kappa_{b}}\hat{b}^{in}(t),\\
&\dot{\hat{m}}_{1}=-i\Delta_{m1}\hat{m}_{1}-ig_1\hat{a}-\kappa_{m1}m_1+\sqrt{2\kappa_{m1}}\hat{m}_1^{in}(t),\\
&\dot{\hat{m}}_{2}=-i\Delta_{m2}\hat{m}_{2}-ig_2\hat{a}-\kappa_{m2}m_2-ig_{mb}\hat{m}_{2}(\hat{b}+\hat{b}^{\dagger})\\&+\Omega_{d}+\sqrt{2\kappa_{m2}}\hat{m}_2^{in}(t).
\end{split} 
\end{equation}
Where $\kappa_{b}$ is the dissipation rate of the phonon mode, and $\hat{b}^{in}(t)$, $\hat{m}_{j}^{in}(t)$ and $\hat{a}^{in}(t)$ are the vacuum input noise operators which have zero mean values i.e.,
 $\langle\hat{q}_{\text{in}}\rangle=0$~\cite{doi:10.1063/1.5027122, aspelmeyer2014cavity}, and ($q=a, m, b$). The magnon mode $m_{2}$ is strongly driven by a microwave drive that causes a large steady-state amplitude $\abs{\langle m_{2s}\rangle}\gg 1$ of magnon mode, and due to beam splitter interaction, this leads to the large steady-state amplitude of the cavity mode $\abs{\langle a_{s}\rangle}\gg 1$. Consequently, we can linearize the quantum Langevin equations around the steady-state values and take only the first-order terms in the fluctuating operator: $\langle\hat{O}\rangle={O}_{s}+\hat{O}_{-}e^{-i\delta t}+\hat{O}_{+}e^{i\delta t}$~\cite{PhysRevA.83.043826}, here $\hat{O}= a, b, m_{j}$. First, we consider the zero-order solution, namely, steady-state solutions which are given by
\begin{equation}\label{eq:steadystate}    
\begin{split}
&a_{s}=-i\sum_{1,2}\frac{g_j m_{js}}{\kappa_{a}+i\Delta_{a}}, \\& b_{s}=\frac{-ig_{mb}\mid{m_{2s}}\mid^{2}}{\kappa_b+i\omega_{b}}, \\ & m_{1s}=\frac{-ig_1 a_s}{\kappa_{m1}+i\Delta_{m1}}, m_{2s}=\frac{\Omega_{d}-ig_2 a_s}{\kappa_{m2}+i\tilde\Delta_{m2}}, \\&
\tilde{\Delta}_{m2}=\Delta_{m2}+g_{mb}(b_{s}+b_{s}^{\star}).
\end{split} 
\end{equation}
We assume that the coupling of the external microwave drive on magnon mode $m_{2}$ is much stronger than the amplitude $\epsilon_{p}$ of the probe field. Under this assumption,  the linearized quantum Langevin equations can be solved by considering the first-order perturbed solutions and ignoring all higher order terms of $\epsilon_{p}$. The solution for the cavity mode is given by
\begin{equation}\label{eq:sol}
a_{-}=\varepsilon_{p}\left[A^{\prime}+C_{1}^{\prime}+\frac{g_{2}^2}{\beta^{\prime}}+\frac{\alpha^{\star}\alpha^{\prime}}{\beta^{\star} \beta^{\prime}+A^{\star}-C_1^{\star}+\frac{g_2^{2}}{\beta^{\star}}}\right]^{-1},
\end{equation}
where
\begin{equation*}    
\begin{split}
&A=\kappa_{a}+i(\Delta_a+\delta), B=\frac{G_{mb}^{2}\omega_{b}}{\omega_{b}^{2}-\delta^{2}+i\delta \kappa_{b}},\\ & C_1=\frac{g_{1}^{2}}{\kappa_{m1}+i(\Delta_{m1}+\delta)}, C_2=\frac{g_{2}^{2}}{\kappa_{m2}+i(\tilde\Delta_{m2}+\delta)},\\& A^{\prime}=\kappa_{a}+i(\Delta_a-\delta), B^{\prime}=\frac{G_{mb}^{2}\omega_{b}}{\omega_{b}^{2}-\delta^{2}-i\delta \kappa_{b}}, \\& C_1^{\prime}=\frac{g_{1}^{2}}{\kappa_{m1}+i(\Delta_{m1}-\delta)}, C_2^{\prime}=\frac{g_{2}^{2}}{\kappa_{m2}+i(\tilde\Delta_{m2}-\delta)},\\&
\alpha=\frac{g_2^{2}B}{C_2+iB}, \alpha^{\prime}=\frac{g_2^{2}B^\prime}{C_2^{\prime}+iB^{\prime}},\\&
\beta=C_{2}-i\frac{C_2^{\prime^\star}B}{C_2^{\prime^\star}+i{B}}, \beta^{\prime}=C_{2}^\prime-i\frac{C_{2}^{\star}B^{\prime}}{C_{2}^{\star}+i{B^\prime}}.
\end{split} 
\end{equation*}
Here $G_{mb}=i\sqrt{2}g_{mb}m_{2s}$ is the effective magnon-phonon coupling. We use the input-output relation for the cavity field $\varepsilon_{out}={\varepsilon}_{in}-2\kappa_{a}\langle a\rangle$~\cite{gardiner2004quantum}, and the amplitude of the output field can be written as
\begin{figure}[ht]
\centering
{\includegraphics[width=4.2cm,height=3.0cm]{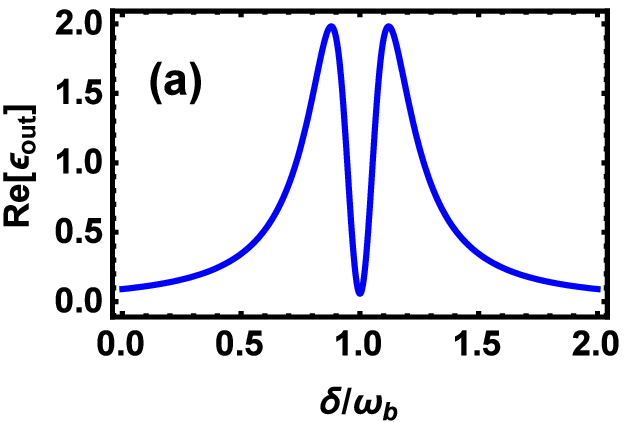}}%
{\includegraphics[width=4.2cm,height=3.0cm]{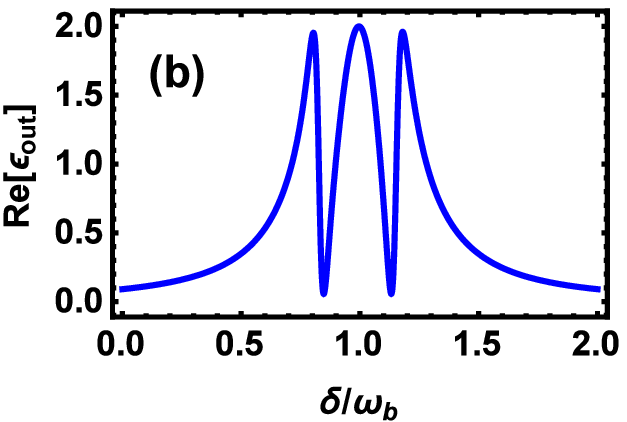}}\\%
{\includegraphics[width=4.2cm,height=3.0cm]{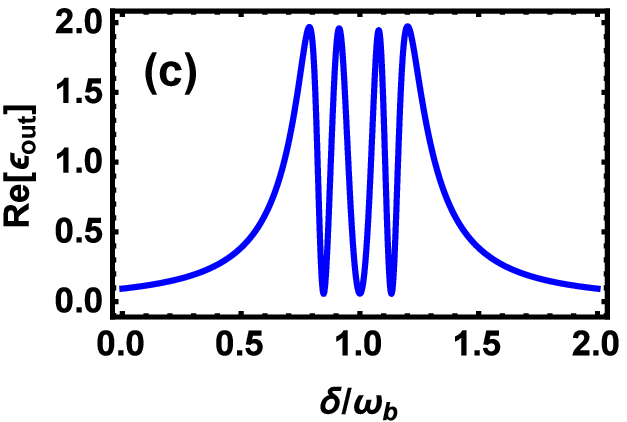}}%
{\includegraphics[width=4.2cm,height=3.0cm]{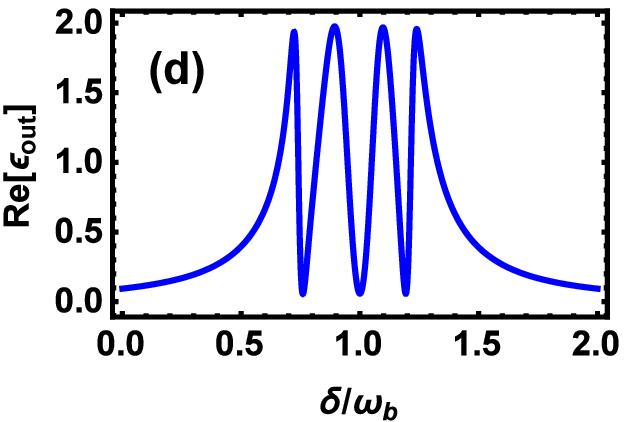}}%
\caption{(Color online).  Absorption Re$[\varepsilon_{out}]$ profiles are shown against the normalized probe field detuning $\delta/\omega_{b}$. (a) $g_{1}=g_{mb}=0$, $g_{2}/2\pi=1.2$ MHz and (b) $g_{1}=0$, $g_{2}/2\pi=1.2$ MHz, $G_{mb}/2\pi=2.0$ MHz (c) $g_{1}/2\pi=g_{2}/2\pi=1.2$ MHz, $G_{mb}/2\pi=2$ MHz and (d) $g_{1}/2\pi=g_{2}/2\pi=1.2$ MHz, $G_{mb}/2\pi=3.5$ MHz. The other parameters are  given in Sec.~\ref{sec:MMITwind}.}\label{fig:2}
\end{figure} 
\begin{equation}\label{eq:output}
\varepsilon_{out}=\frac{2\kappa_a a_{-}}{\varepsilon_p}.
\end{equation}
The real and imaginary parts of $\varepsilon_{out}$ account for in-phase (absorption) and out of phase (dispersion) output field quadratures at probe frequency. 
\section{MMIT windows profile}\label{sec:MMITwind}
For the numerical calculation, we use parameters from a recent experiment on a hybrid magnomechanical system~\cite{Zhange1501286}, unless stated differently.
Frequency of the cavity field $\omega_{a}/2\pi=10$ GHz,  $\omega_{b}/2\pi=10$ MHz, $\kappa_{b}/2\pi=100$ Hz, $\omega_{1,2}/2\pi=10$ GHz, $\kappa_{a}/2\pi=2.1$ MHz, $\kappa_{m1}/2\pi=\kappa_{m2}/2\pi=0.1$ MHz, $g_{1}/2\pi=g_{2}/2\pi =1.5$ MHz, $G_{mb}/2\pi=3.5$ MHz, $\Delta_a=\omega_{b}$,  $\Delta_{mj}=\omega_{b}$, $\omega_{d}/2\pi=10$ GHz.

We first illustrate the physics behind the multiband transparency by systematically investigating the role of different couplings in the model.
Fig.~2 displays the response of the probe field in the absorption spectrum of the output field for different coupling strengths. In Fig.~2(a), we assume the magnon-phonon coupling ($g_{mb}$) and  magnon mode $m_{1}$ coupling ($g_{1}$) with the cavity are absent. Therefore, only magnon mode $m_{2}$ is coupled with the cavity. Under these considerations, we observe a magnon induced transparency (MIT) in which a typical Lorentzian peak of the output spectrum of the simple cavity splits into two peaks with a single dip, as shown in Fig.~2(a). The width of this transparency window can be controlled via microwave driving field power and the magnon-photon coupling $g_{2}$. On increasing the coupling strength $g_{2}$ the width of the window increases, and vice versa.
\begin{figure}[ht]
\centering
{\includegraphics[width=4.2cm,height=3.0cm]{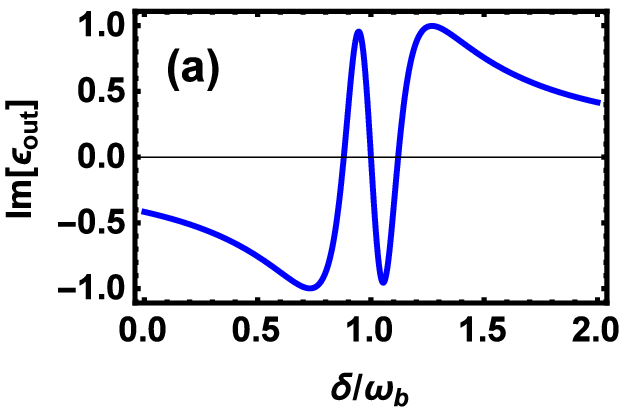}}%
{\includegraphics[width=4.2cm,height=3.0cm]{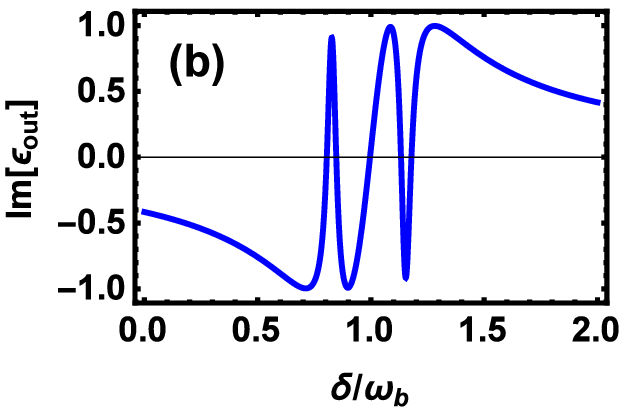}}\\%
{\includegraphics[width=4.2cm,height=3.0cm]{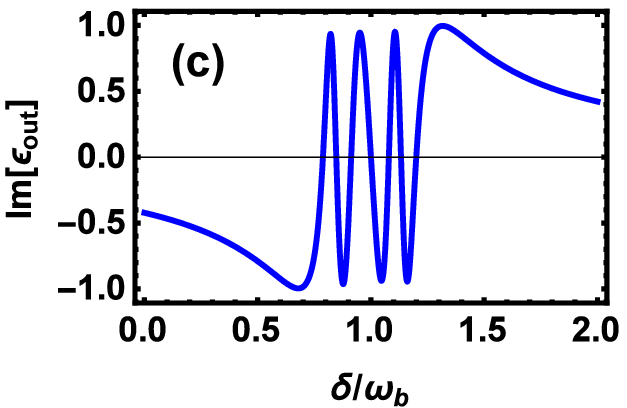}}%
{\includegraphics[width=4.2cm,height=3.0cm]{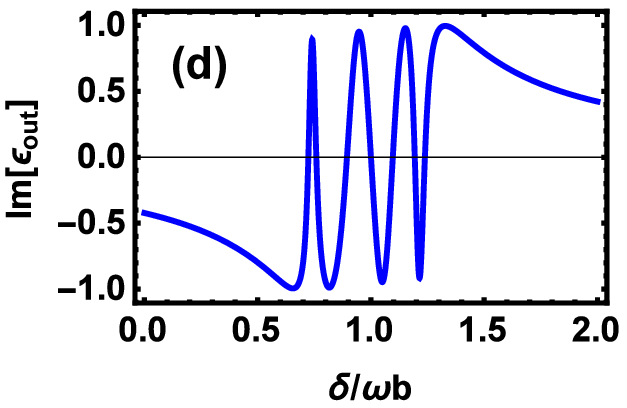}}%
\caption{(Color online) Dispersion Im$[\varepsilon_{out}]$ profiles are shown against the normalized probe detuning $\delta/\omega_{b}$. (a) $g_{1}=g_{mb}=0$ and $g_{2}/2\pi=1.2$ MHz and (b) $g_{1}=0$, $g_{2}/2\pi=1.2$ MHz, $G_{mb}/2\pi=2.0$ MHz (c), (d) $g_{1}/2\pi=g_{2}/2\pi=1.2$ MHz, and (c) $G_{mb}/2\pi=2$ MHz and (d) $G_{mb}/2\pi=3.5$ MHz. The other parameters are given in Sec.~\ref{sec:MMITwind}.}
\end{figure}

We observe two transparency windows in the absorption as we switch on the magnon-phonon coupling ($g_{mb}$) and keeping $g_{1}=0$. Due to the non-zero magnetostrictive interaction, single MIT window in Fig.~2(a) splits into double window shown in Fig.~2(c). The right transparency window in Fig.~2(c) is associated with magnon-phonon interaction, and this is so called magnomechanically induced transparency (MMIT)~\cite{Zhange1501286} window.
We can observe double MIT by removing magnon-phonon coupling $g_{mb}$, and considering non-zero couplings between the magnon modes and the cavity field. 

Finally, if we consider all three couplings simultaneously non-zero, then the transparency window splits into three windows consist of four peaks and three dips, this is shown in Fig.~2(c). 
In this case, one window is associated with the magnomechanical interaction, and the rest of the two are induced by magnon-photon couplings.
The width and peaks separation of these windows increases and broadens, respectively, at higher values of magnon-phonon coupling $G_{mb}$, which can be seen in Fig.~2(d). Moreover, we have a symmetric multi-window transparency profile where the splitting of the peaks occurs at side-mode frequencies $\omega_{p}=\omega_{b}\pm\omega_{d}$. 

In Figs.~3(a-d), we plot the dispersion spectrum of the output field versus normalized frequency of the probe field. The single MIT dispersion spectrum in the absence of YIG1 and magnon-phonon coupling $g_{mb}$ is shown in Fig.~3(a). 
The dispersion spectra for the case of $g_{1}=0$, $g_{2}\neq 0$ and $g_{mb}\neq 0$ is plotted in the Fig.~3(b). In the presence of all three couplings, the dispersion spectrum of the output field is given in the Figs.~3(c-d). It is clear from Figs.~3(c-d), by the increase in the effective magnon-phonon coupling $G_{mb}$, the transparency windows become wider. We like to point out that the magnomechanically induced amplification (MMIA) of the output field, in our system, can be obtained in the blue detuned regime; $\Delta_{m2}=-\omega_{b}$. 
\begin{figure}[ht]
\centering
{\includegraphics[width=4.2cm,height=3.0cm]{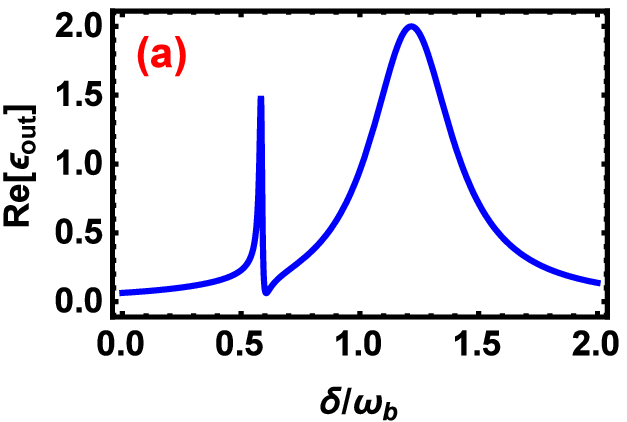}}%
{\includegraphics[width=4.2cm,height=3.0cm]{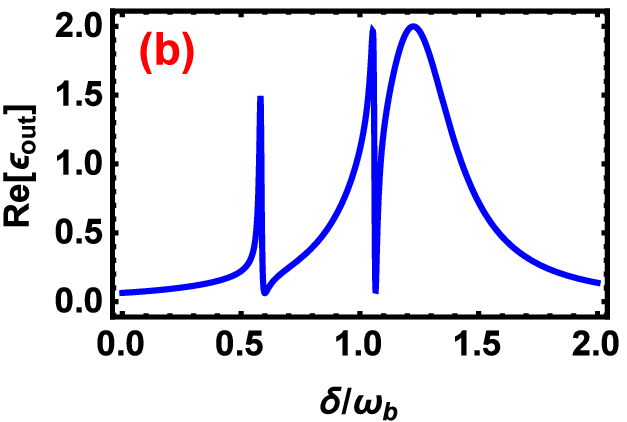}}\\%
{\includegraphics[width=4.2cm,height=3.0cm]{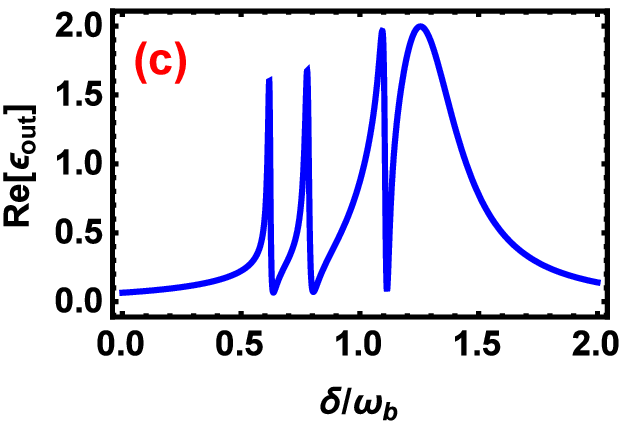}}%
{\includegraphics[width=4.2cm,height=3.0cm]{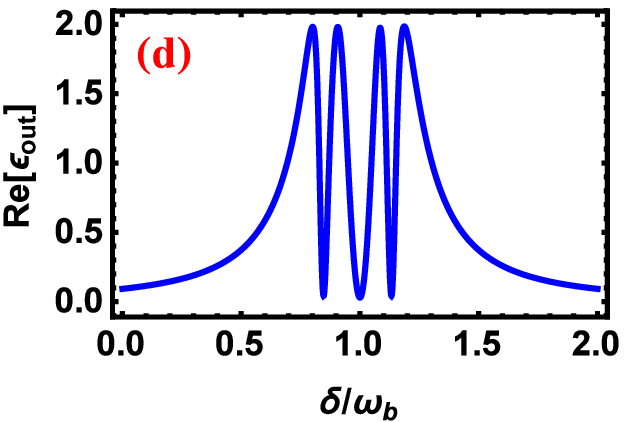}}%
\caption{(Color online) Fano line shapes in the asymmetric absorption Re$[\varepsilon_{out}]$ profiles  are shown against the normalized probe frequency $\delta/\omega_{b}$. (a) $\Delta_{m2}=0.7\omega_{b}$, $g_{2}=1.5$ MHz, $g_{1}=g_{mb} =0$, and (b) $\Delta_{m2}=0.7\omega_{b}$, $g_{1}=0$, $g_{2}=1.5$ MHz, $G_{mb}=3.5$ MHz. (c) $\Delta_{m1,2}=0.7\omega_{b}$, $g_{1}=g_{2}/2\pi=1.5$ MHz and $G_{mb}/2\pi=3.5$MHz, and (d) $\Delta_{m1,2}=\omega_{b}$ , $g_{1}=g_{2}/2\pi=1.5$ MHz and $G_{mb}/2\pi= 3.5$ MHz. In all panels, $g_{1}=g_{2}/2\pi=1.5$ MHz, $G_{mb}/2\pi=3.5$ MHz, and rest of the parameters are give in Sec.~\ref{sec:MMITwind}.}
\end{figure} 
\section{Fano resonances in the output field}\label{sec:fano}

In the following, we discuss the emergence and physical mechanism of the Fano line shapes in the output spectrum. The shape of the Fano resonance is distinctly different than the symmetric resonance curves in the EIT, MIT, optomechanically induced transparency (OMIT) and MMIT windows~\cite{PhysRevA.87.063813, Zhange1501286}. Fano resonance has observed in the systems in which EIT has reported by a suitable selection of the system parameters~\cite{Zhange1501286, PhysRevA.87.063813, PhysRevA.97.033812, refId0, yasir2016controlled, Akram_2015, zhang2017optical}. The physical origin of Fano resonance in the systems having optomechanical-like interactions has explained due to the presence of non-resonant interactions. For example, in a standard optomechanical system, if the anti-Stokes process is not resonant with the cavity frequency, asymmetric Fano shapes appear in the spectrum~\cite{PhysRevA.87.063813, PhysRevA.97.033812, refId0}. In our system, this corresponds to $\Delta_{m1}\neq\omega_{b}$, because instead of a cavity mode, magnon mode $m_{1}$ is coupled with phonon mode via optomechanical-like interaction. 
\begin{figure}[ht]
	\centering
	\includegraphics[width=4.5cm,height=3.0cm]{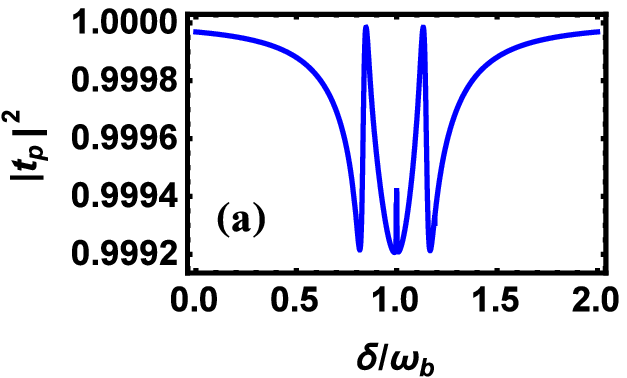}%
	\includegraphics[width=4.5cm,height=3.0cm]{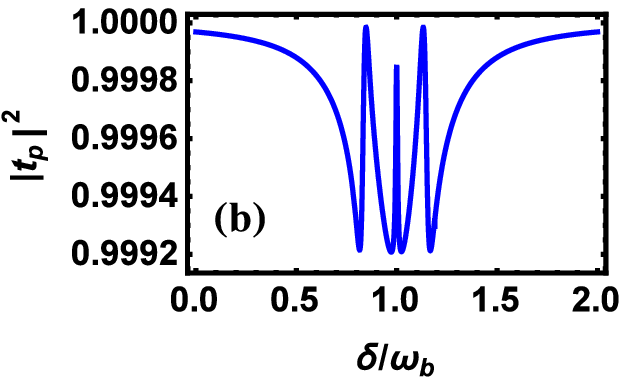}\\%
	\includegraphics[width=4.5cm,height=3.0cm]{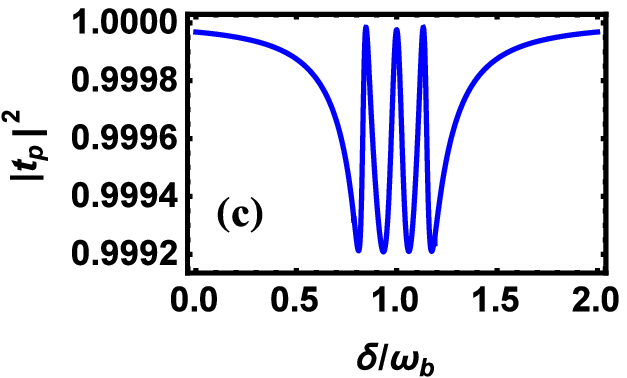}%
	\includegraphics[width=4.5cm,height=3.0cm]{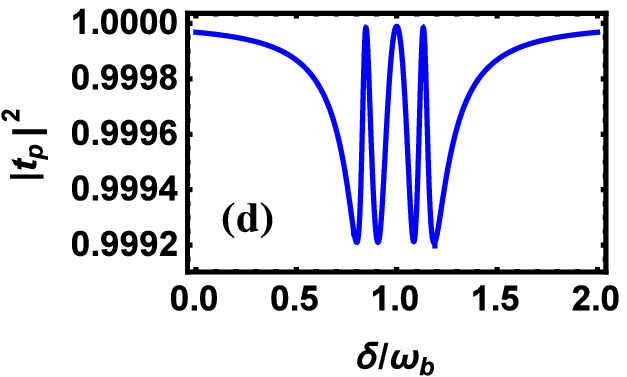}%
	\caption{(Color online). The transmission $|t_{p}|^{2}$ spectrum as a function of normalized probe field frequency $\delta/\omega_{b}$ is shown for different values of $g_{1}$. (a) $g_{1}/2\pi=0.5$ MHz (b) $g_{1}/2\pi=0.8$ MHz (c) $g_{1}/2\pi=1.2$ MHz (d) $g_{1}/2\pi=1.5$ MHz. In all panels, $g_{2}/2\pi=1.5$ MHz, $G_{mb}/2\pi=3.5$ MHz and the other parameters are given in Sec.~\ref{sec:MMITwind}.}
\end{figure}
The asymmetric Fano shapes can be seen in Figs.~4(a-c) for different non-resonant cases, where  the absorption spectrum of the output field as a function of normalized detuning $\delta/\omega_{b}$ is shown.  In Fig.~4(a), we consider $g_{1}=g_{mb}=0$, and coupling of the magnon mode $m_{2}$ with the cavity is non-zero. Due to the presence of non-resonant process ($\Delta_{m2}=0.7\omega_{m}$), the absorption spectrum of the symmetric MIT (Fig.~2(a)) profile changes into asymmetric window profile, as shown in Fig.~4(a).  
Such asymmetric MIT band can be related to Fano-like resonance, emerging frequently in optomechanical systems~\cite{PhysRevA.87.063813, PhysRevA.97.033812,  refId0, yasir2016controlled, Akram_2015}. If we remove YIG1 and consider only YIG2 is coupled with the cavity mode, and $\Delta_{m2}=0.7\omega_{m}$. We observe double Fano resonance in the output spectrum, which is shown in Fig.~4(b). Similarly, in the presence of all three couplings and $\Delta_{m1, m2}=0.7\omega_{m}$, the double Fano resonance goes over to a triple Fano profile, as shown in Fig.~4(c). This is because the cavity field can be build up by three coherent routes provided by the three coupled systems (the magnons, cavity, and phonon modes), and that can interfere with each other. The Fano resonances disappear when we consider a resonant case $\Delta_{m1}=\Delta_{m2}=\omega_{b}$, as shown in Fig.~4(d).

\section{Numerical results for slow and fast lights}\label{sec:SlowFast}

Here we investigate the transmission and group delay of the output signal, and show the effect of the magnon-photon and
magnon-phonon couplings on the transmission spectrum. From Eq.~\eqref{eq:output}, the rescaled transmission field corresponding to the probe field can be expressed as
\begin{equation}
t_{p}=\frac{\varepsilon_p-2\kappa_a a_{-}}{\varepsilon_p}. 
\end{equation}
In Figs.~5(a-d), we plot the transmission spectrum of the probe field against the scaled detuning $\delta/\omega_{b}$, for different values of $g_{1}$. It is clear from Fig.~5(a), the transmission peak  associated with the magnon-photon coupling of YIG1 is smaller than the other two peaks. This is because in Fig.~5(a) $g_{1}$ coupling is weaker than the other two interactions $g_{2}$ and $G_{mb}$ present in the system. 
By increasing the coupling strength $g_{1}$, the peak of the middle transparency profile grows up in height and reaches close to unity, as shown in Figs.~5(b-c). In addition, Fig.~5(d) shows that the width of the transparency window can be increased at higher higher values of the magnon-photon coupling $g_{1}$.

\begin{figure}[ht]
	\centering
	\includegraphics[width=4.5cm,height=3.2cm]{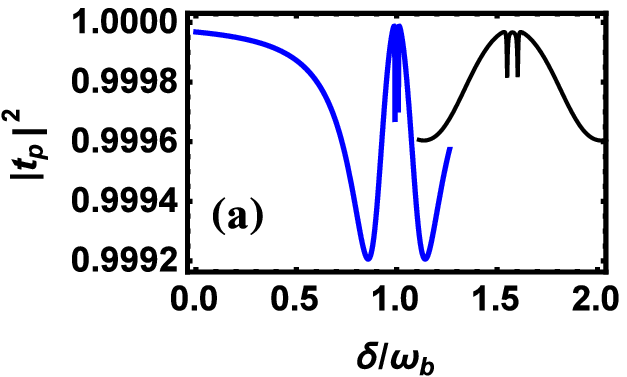}%
	\includegraphics[width=4.5cm,height=3.2cm]{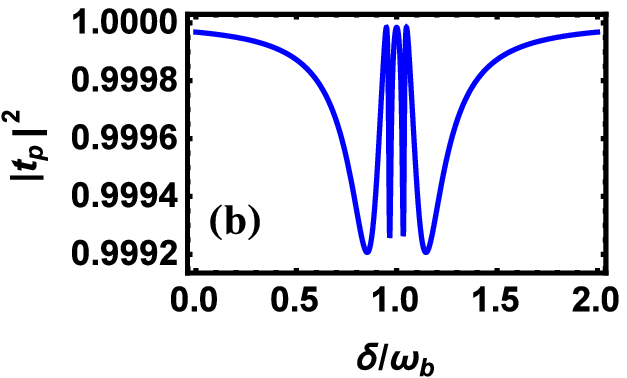}\\%
	\includegraphics[width=4.5cm,height=3.2cm]{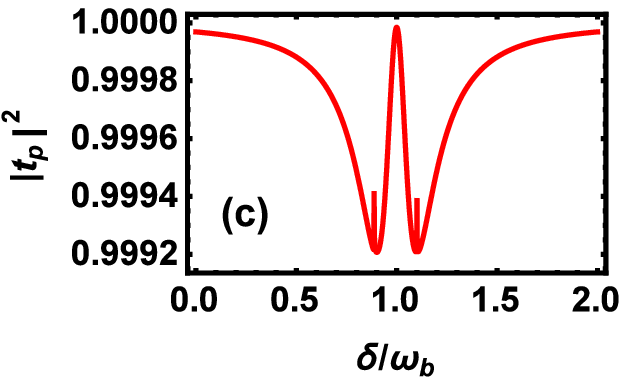}%
	\includegraphics[width=4.5cm,height=3.2cm]{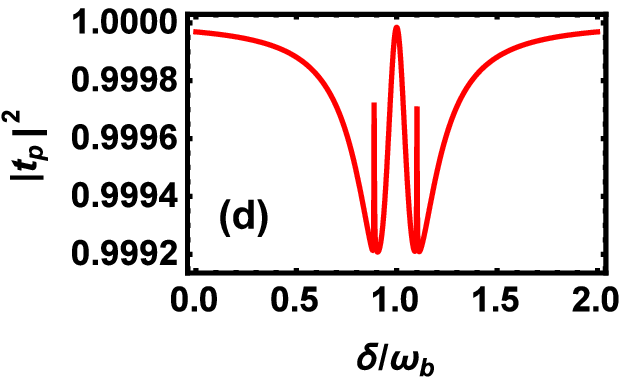}%
	\caption{(Color online). The transmission $|t_{p}|^{2}$ spectrum as a function of normalized probe field frequency $\delta/\omega_{b}$ is shown. (a) $G_{mb}/2\pi=0.5$ MHz (b) $G_{mb}/2\pi=1.0$ MHz. In (c) $g_{2}/2\pi=0.4$ MHz, and (d) $g_{2}/2\pi=0.8$ MHz. The other parameters are same as in Fig.~5.}
\end{figure}
In Figs.~6(a-b), the transmission spectrum of the probe field as a function of dimensionless detuning is shown for different values of $G_{mb}$. In Figs.~6(a-b), we consider both $g_{1}$ and $g_{2}$ to be the same in the strong coupling regime. However, the effective coupling $\tilde{g}_{2}=g_{2}\alpha_{s}$ depends on the steady-state amplitude of the cavity field $\alpha_{s}$ which depends on the $m_{2s}$. Consequently, $\tilde{g}_{2}$ and $G_{mb}$ are related and it can be seen from Eq.~\eqref{eq:steadystate}. For a smaller value of $G_{mb}$ in Fig.~6(a), we have two small peaks associated with $g_{2}$ and $G_{mb}$, in addition, the third-highest peak is associated with $g_{1}$. For a fixed value of $g_{mb}$, if we increase $G_{mb}$, it increases $\tilde{g}_{1}$, and the peaks associated with these two couplings become more visible, as shown in Fig.~6(b).
Similarly, in Fig.~6(c-d), we observe a similar increase in the height of two peaks associated with $g_{2}$ and $G_{mb}$, for the variation in $g_{2}$. 

The phase $\phi_{t}$ of the transmitted probe field $t_{p}$ is given by the relation $\phi_{t}= \text{Arg}[t_{p}]$. The plot of $\phi_{t}$ as a function of normalized detuning $\delta/\omega_{b}$ is shown in Fig.~7. In the inset of Fig.~7(a), we consider both $g_{1}$ and $g_{mb}$ are switched off, and only $g_{2}$ is non-zero. 
This gives a conventional phase of the transmitted field with a single MIT curve, which appears similar to the standard single OMIT curve~\cite{PhysRevA.87.063813}. 
In Fig.~7(b), we switch-off the YIG1 coupling with the field ($g_{1}=0$), and the other two couplings are present ($g_{mb}\neq 0, g_{2}\neq 0$), due to which the single transparency window splits into a double window. If we keep all three couplings non-zero, we get triple transparency window which is shown in Fig.~7(c).  
\begin{figure}[ht]
	\centering
	\includegraphics[width=4.2cm,height=3cm]{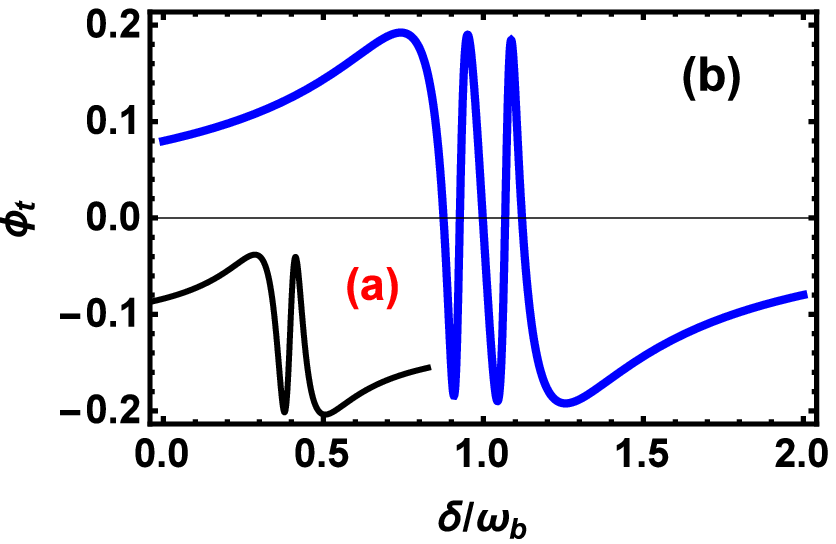}%
	\includegraphics[width=4.3cm,height=3.0cm]{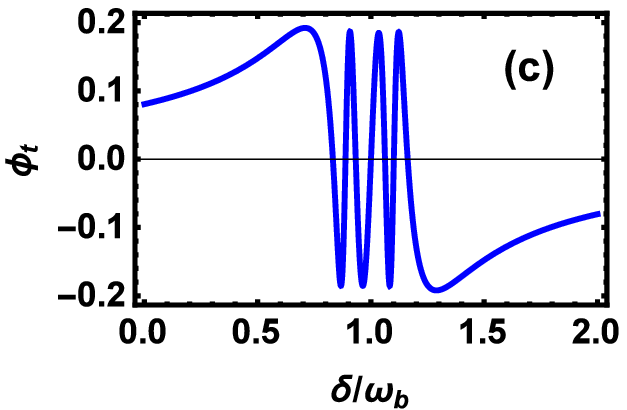}%
	\caption{(Color online) The phase $\phi_{t}$ of the transmitted probe field versus normalized detuning $\delta/\omega_{b}$ for different coupling strengths. (a) $g_{1}=g_{mb}=0$, (b) $g_{1}=0$, $g_{2}/2\pi=1.5$ MHz, $G_{mb}/2\pi=4$ MHz (c) $g_{1}/2\pi=g_{2}/2\pi=1.5$ MHz, and $G_{mb}/2\pi=4$ MHz. Rest of parameters are given in Sec.~III.}
\end{figure} 
\begin{figure}[ht]
\centering
\includegraphics[width=4.4cm,height=3.1cm]{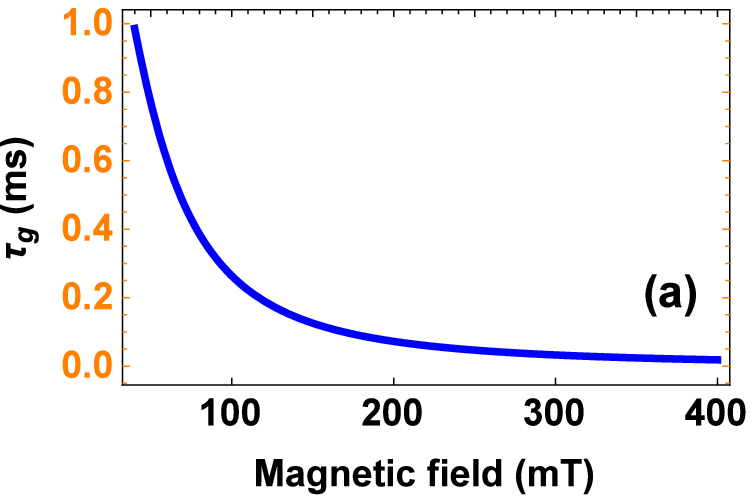}%
\includegraphics[width=4.4cm,height=3.1cm]{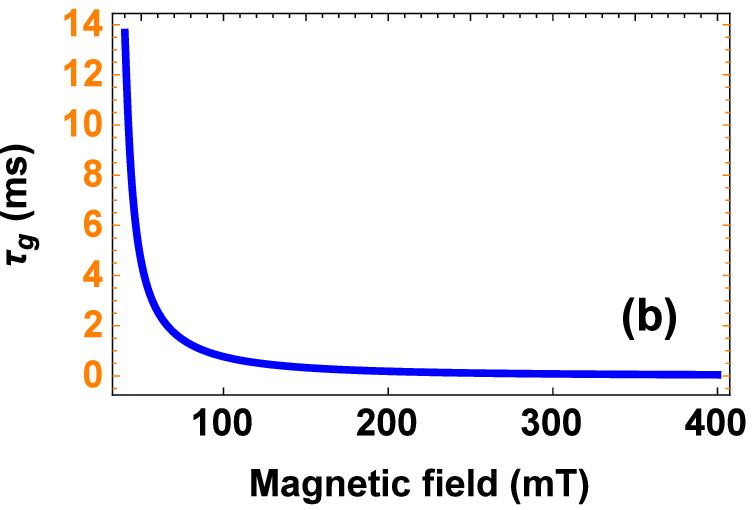}%
\caption{(Color online) Group delay $\tau_{g}$ of the output probe field against the amplitude of the magnetic field $B_{0}$ for (a) $g_{1}=0$, and (b) $g_{1}/2\pi=1.5$ MHz.
The other parameter are $g_{2}/2\pi=1.5$, $G_{mb}/2\pi=3.5$ MHz, $\kappa_{b}/2\pi=100$ Hz, $\kappa_{m1}/2\pi=\kappa_{m2}/2\pi=0.1$ MHz, $\kappa_{a}/2\pi=2.1$ MHz and $\Omega_{d}=1.2$ THz.}
\end{figure} 

The transmitted probe field phase is associated with the group delay $\tau_{g}$ of the output field and it is defined as 
\begin{equation}\label{eq:grdelay}
\tau_{g} = \frac{\partial\phi(\omega_{p})}{\partial\omega_{p}},
\end{equation} 
which means a more rapid phase dispersion leads to a larger group delays and vice versa.
In addition, a negative slope of the phase represents a negative group delay or fast light ($\tau_{g}<0$) whereas, a positive slope of the transmitted field indicates positive group delay or slow light ($\tau_{g}>0$). From Fig.~7, we observe that in the regime of the narrow transparency window, there is a rapid variation in the probe phase, and this rapid phase dispersion can lead to a significant group delay.

Fig.~8 shows the group delay $\tau_{g}$ can be tuned by the variation of the bias magnetic field $B_{0}$ applied on YIG2. In the absence of YIG1 (Fig.~8(a)), we have a lower slope of Eq.~\eqref{eq:grdelay}, as a result, a maximum group delay of $\tau_{g}=1$ ms is achieved. This group delay can be enhanced by one order of magnitude once second YIG is introduced see Fig.~8(b). The slope of Fig.~8(b) become steeper and the time delay for slow light is increased up to 13.8 ms. This shows the two YIGs system is a good choice to observe a longer group delay in a magnomechanical system while a single YIG system cannot do so. Moreover, the numerical value of the group delay $\tau_{g}$ can be tuned from positive (slow light) to negative (fast light) by tuning the magnon field detuning $\Delta_{m1}=\omega_{b}$ to $\Delta_{m1}=-\omega_{b}$.
Here, it is worth mentioning that Fig.~8(b) can be switched into fast light with a maximum group delay in the order of $\tau_{g}\approx -1.4$ ms in the presence of both YIGs and we not show in the figure. This negative group delay for the fast light propagation is one order of magnitude greater than a single YIG magnomechanical system~\cite{Kong:19}.

 \begin{figure}[ht]
\centering
\includegraphics[width=4.4cm,height=3.1cm]{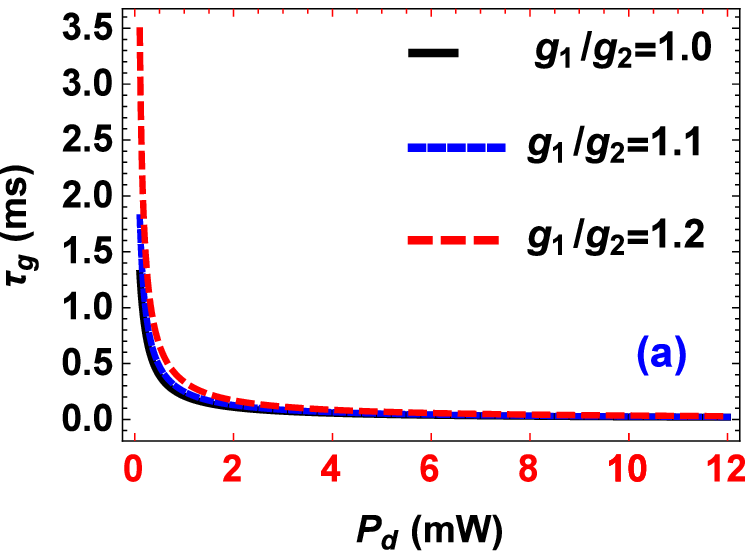}%
\includegraphics[width=4.4cm,height=3.1cm]{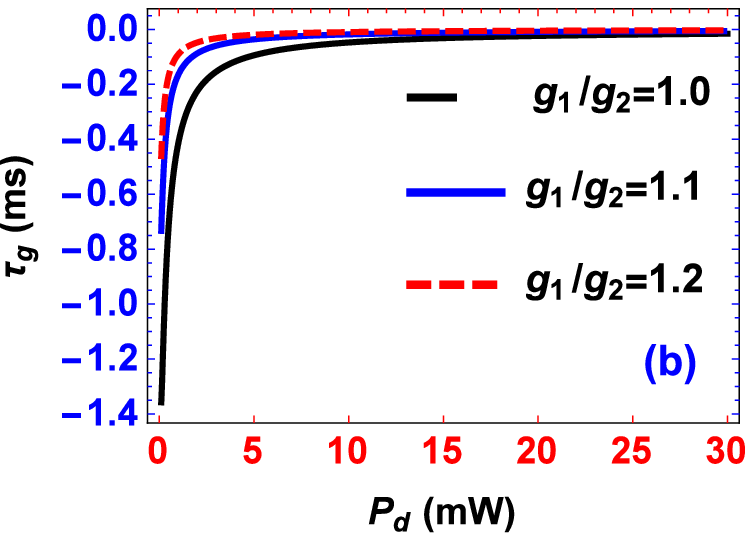}%
\caption{(Color online) The group delay $\tau_{g}$ of the transmitted probe field as a function of the driving power $P_{d}$ for several values of the magnon-photon couplings. (a) $\Delta_{m1}=\omega_{b}$, (b) $\Delta_{m1}=-\omega_{b}$, and the other parameter are same as given in Fig.~8.}
\end{figure}
 
Finally, we investigate the control of group delay with the external microwave driving power and magnon-photon couplings. 
For this purpose, in Figs.~9(a-b), we plot $\tau_g$ against the driving power for different strengths of the magnon-photon coupling of YIG1 with respect to the coupling frequency of YIG2. 
Fig.~9(a) shows that the magnitude of the group delay increases with the increase of $g_{1}$ corresponding to $g_{2}$, which indicates an enhanced group delay of the transmitted probe field in a two-YIG system. We tuned the coupling strength of YIG1 ($g_{1}$) for different values via keeping the coupling strength of YIG2 ($g_{2}$) constant. This shows increasing the magnon-photon coupling strength increases the group delay of the transmitted probe field and vice versa. This helps us to obtain larger group delays at relatively weak magnon-photon coupling strengths which is not otherwise possible with a single YIG magnomechanical system~\cite{Kong:19}. 
Similar results can also be obtained by increasing the magnon-photon coupling $g_{2}$ and fixing $g_{1}$.
For the blue detuned regime $\Delta_{m1}=-\omega_{b}$, group delay becomes negative. However, the effect of magnon-photon couplings remains the same, as shown in Fig. 9(b).
From Fig.~8 and Fig.~9, we see that two YIGs magnomechanical system provides not only extra tunability, but also drastically enhances the group delays compared to single YIG system studied in Ref.~\cite{Kong:19}. Our system can be used as a tunable switch, which can be controlled via different system parameters, and our results are comparable with the existing proposals based on various hybrid quantum systems~\cite{PhysRevA.77.050307, Ullah_2019, PhysRevA.87.013824, doi:10.1063/1.5089435}.

\section{Conclusion}\label{sec:concul}
We have investigated the transmission and absorption spectrum of a weak probe field under a strong control field in a hybrid magnomechanical system in the microwave regime. Due to the presence of a nonlinear phonon-magnon interaction, we observed magnomechanically induced transparency (MMIT), and the photon-magnon interactions lead to magnon induced transparency (MIT). We found single MMIT, a result of the single-phonon process, and found two MIT windows in the output probe spectra due to the presence of two magnon-photon interactions. This is demonstrated by plotting the absorption, dispersion, and transmission of the output field. We discussed the emergence of Fano resonances in the output field spectrum of the probe field. These asymmetric line shapes appeared due to the presence of anti-Stokes processes in the system. We examined conditions of slow and fast light propagation in our system, which can be controlled by different system parameters. It has shown that in a two YIGs magnomechanical system, the tunability of the first coupling strength (YIG1) corresponding to the coupling of the second YIG (YIG2) has an immense effect on the slow and fast light and vice versa. This not only helped to investigate larger group delays at a weak magnon-photon coupling but also enhanced the group delay of the transmitted probe field, which is not possible in a single YIG system. Our results suggest that this system may find its applications to implement multi-band quantum memories~\cite{8701447}.

\section*{Acknowledgment} We acknowledge Prof. M. Cengiz Onbasli for fruitful discussions. We also thank the anonymous referees for their valuable comments which greatly improved our manuscript.

\bibliographystyle{apsrev4-1}
\bibliography{bibliography,revtex-custom}
  
\end{document}